\documentclass[12pt,preprint]{aastex}

\usepackage{epsfig,graphicx,amssymb,epstopdf,mathrsfs}

\newcommand{\enu}{\varepsilon_\nu}

\newcommand{\cerenkov}{\v{C}erenkov}

\def\arcdeg{\hbox{$^\circ$}}
\def\arcmin{\hbox{$^\prime$}}

\def\simlt{\mathrel{\hbox{\rlap{\hbox{\lower4pt\hbox{$\sim$}}}\hbox{$<$}}}}
\def\simgt{\mathrel{\hbox{\rlap{\hbox{\lower4pt\hbox{$\sim$}}}\hbox{$>$}}}}

\newcommand{\xray}{\mbox{X-ray}}

\newcommand{\percmsq}{cm$^{-2}$}
\newcommand{\percmcube}{cm$^{-3}$}
\newcommand{\tnm}[1]{\tablenotemark{#1}}

\newcommand{\perval}[2]{{#1\mbox{$^{#2}$}}}

\newcommand{\persec}{\perval{\rm s}{-1}\/}
\newcommand{\percm}{\mbox{$\cm^{-2}$}}

\newcommand{\pertev}{\perval{\rm TeV}{-1}\/}
\newcommand{\persr}{\perval{\rm sr}{-1}\/}

\newcommand{\erg}{\mbox{$\rm\,erg$}\/}
\newcommand{\cm}{\mbox{$\rm\,cm$}}

\newcommand{\cgsflux}{\erg\,\percm\,\persec}
\newcommand{\cgsfluxsr}{\erg\,\percm\,\persec\,\persr}

\begin{document}

\title{Sub-PeV Neutrinos from TeV Unidentified Sources in the Galaxy}

\author{D.~B. Fox\altaffilmark{1}, 
        K. Kashiyama\altaffilmark{1},
        P. M\'eszar\'os\altaffilmark{1}}

\email{dfox@astro.psu.edu, kzk15@psu.edu, nnp@astro.psu.edu}

\altaffiltext{1}{Department of Astronomy \& Astrophysics; Department
  of Physics; Center for Particle \& Gravitational Astrophysics;
  Pennsylvania State University, University Park, PA 16802, USA}

\slugcomment{Submitted to ApJ}
\shorttitle{Sub-PeV Neutrinos from TeV UnID Sources}
\shortauthors{Fox, Kashiyama \& M\'eszar\'os}


\begin{abstract}
The IceCube collaboration discovery of 28 high-energy neutrinos over
the energy range $\mbox{30 TeV}\simlt \enu \simlt \mbox{1 PeV}$, a
4.3$\sigma$ excess over expected backgrounds, represents the first
high-confidence detection of cosmic neutrinos at these energies.
In light of this discovery, we explore the possibility that some of
the Sub-PeV cosmic neutrinos might originate in our Galaxy's TeV
unidentified (TeV UnID) sources. While typically resolved at TeV
energies, these sources lack prominent radio or \xray\ counterparts,
and so have been considered promising sites for hadron acceleration
within our Galaxy.
Modeling the TeV UnID sources as Galactic hypernova remnants, we
predict Sub-PeV neutrino fluxes and spectra consistent with their
contributing a minority of $n_\nu\simlt 2$ of the observed events.
This is consistent with our analysis of the spatial
distribution of the Sub-PeV neutrinos and TeV UnID sources, which
finds that a best-fit of one, and maximum of 3.8 (at 90\%-confidence),
of the $\approx$16 non-atmospheric sub-PeV neutrinos may originate in
TeV UnID sources, with the remaining 75\% to 95\% of events being
drawn from an isotropic background.
If our scenario is correct, we expect excess Sub-PeV neutrinos to
accumulate along the Galactic plane, within $|\ell| \simlt 30\arcdeg$
of the Galactic center and in the Cygnus region, as observations by
IceCube and other high-energy neutrino facilities go forward. Our
scenario also has implications for radio, \xray, and TeV observations
of the TeV UnID sources.
\end{abstract}

\keywords{neutrinos --- %
          gamma rays: general --- %
          cosmic rays --- %
          ISM: supernova remnants}

\maketitle


\section{Introduction}
\label{sec:intro}

The IceCube collaboration recently reported their detection of 28
neutrinos with energies in the range $\mbox{30 TeV} \simlt \enu \simlt
\mbox{1 PeV}$, inconsistent with the expected atmospheric background
of $\approx$12 such events at 4.3$\sigma$ confidence, and indicating
an origin for $\approx$60\% of these events in cosmic sources
(\citealt{Ic3+13-subpev}; see also \citealt{Ic3+13-pev}). The spectrum
of these Sub-PeV cosmic neutrinos can be adequately described by a
power law with particle index $\Gamma \approx 2$ extending up to a
cutoff energy of roughly $\varepsilon_{\rm \nu, max} \approx
1.5$~PeV\footnote{Although see recent alternative analyses suggesting
  consistency with an unbroken power-law spectrum
  \citep{Anchordoqui+13}, or with a two-component spectrum with a
  cutoff at $\varepsilon_{\rm \nu, max} \approx 250$~TeV
  \citep{He+13}.}; the observed surface brightness is $\enu
F_{\varepsilon_{\nu}} \sim 3 \times 10^{-11}$\,\cgsfluxsr.

The sky distribution of the Sub-PeV neutrinos, which include seven
well-localized ($\sim$2\arcdeg\ uncertainty) track-type events and 21
less well-localized ($\sim$15\arcdeg\ uncertainty) cascade-type
events, rules out attributing more than a minority of the cosmic
neutrinos to a single source (e.g., no two track-type events have
overlapping localizations). In a preliminary analysis, the IceCube
team find that the distribution is consistent with isotropy given
their backgrounds and sky coverage \citep{Ic3+13-spatial}. However,
their search for low-probability event clusters reveals that the
single highest-likelihood source position on the sky, with $p=8\%$
after trials, is located roughly 15\arcdeg\ from the Galactic Center,
the result of a cluster of 2 to 7 cascade-type events in this
vicinity. Since this location has a natural significance for Galactic
source populations, and since a population of multiple sources might
well extend over multiple degrees along the Galactic plane, the
IceCube point-source $p$ value of 8\% can be considered conservative
with respect to appropriate, astrophysically-motivated Galactic source
scenarios. Given the near-isotropy of the overall event distribution,
any such Galactic source population will likely account for a minority
of the observed Sub-PeV neutrinos.

In this paper, we explore a single such scenario, first put forward by
\citet{Ioka+10unid}. \S2 introduces the class of TeV gamma-ray
unidentified (TeV UnID) sources and our preferred astrophysical model
for these sources, as relatively old ($T\sim 10^5$\,yr) remnants of
Galactic hypernovae. We calculate the flux and spectrum of Sub-PeV
neutrinos expected from the TeV UnID sources, individually and as a
group, and show consistency with IceCube observations. In \S3 we
explore the expected sky distribution of the Sub-PeV neutrinos from
TeV UnID sources, and consider how many of the observed Sub-PeV events
can be attributed to this population. In \S4 we review our results,
suggest future tests of our hypothesis, and consider the implications
of our findings for high-energy neutrino astronomy and high-energy
astrophysics generally.


\section{TeV Unidentified Sources as Hadron Accelerators}
\label{sec:tevunid}

In order to produce high-energy neutrinos, Galactic sources have to be
efficient cosmic ray accelerators. Supernova remnants are thought to
accelerate protons up to at most $\sim$PeV energies
\citep{Gaisser90book}, resulting in neutrinos of energies
$\enu\lesssim 0.1$\,PeV. Thus, a more promising Galactic source
population at high energies is the hypernova remnants, which can
accelerate protons up to significantly higher energies
\citep{Wang+07crhn,Budnik+08hn} by virtue of their systematically
greater explosion energies and higher shock velocities. Furthermore,
hypernova remnants are likely to have relatively fainter lower-energy
photon emission at late stages, making them harder to detect via
\xray\ and radio observations, and favoring hadronic emission
mechanisms. This set of properties makes them good candidates for the
class of TeV unidentified (TeV UnID) gamma-ray sources
\citep{Ioka+10unid}.

To date, 24 such TeV UnID sources have been found within and near the
Galactic plane\footnote{See Table~\ref{tab:unid} and the TeVCat online
  catalog of TeV gamma-ray sources at
  \url{http://tevcat.uchicago.edu/}} \citep{krp13}; the resolved
nature of these sources and their distribution along the plane, with a
concentration toward the Galactic center, serve to indicate their
Galactic origin.  Observed fluxes of the TeV UnID sources are
typically $\varepsilon_{\gamma} F_{\varepsilon_\gamma} \sim 10^{-12}$
to $10^{-11}$\,\cgsflux\ at $\varepsilon_{\gamma} \sim 1$\,TeV. They
show a power-law spectrum with photon index $\Gamma = 2.1$ to 2.5 at
TeV energies, and exhibit a typical angular size of $\theta\sim
0.05\arcdeg$ to 0.3\arcdeg.  They are classified as TeV UnID sources
on the basis of strong constraints on their lower-energy (\xray\ and
radio) electromagnetic fluxes, e.g., $F_{\rm TeV}/F_{\rm X} \gtrsim
50$ \citep{mub+07,bkh+07,tlh+08} and $F_{\rm TeV}/F_{\rm radio}
\gtrsim 10^3$ \citep{abk06,tlh+08}.  Absence of bright \xray\ and
radio counterparts disfavors leptonic models, since these require an
intense low-energy photon field to be up-scattered into the TeV by
inverse-Compton processes. Instead, hadronic models have been invoked
for these sources, as we consider here.

First, we show that the flux and spectra of Sub-PeV neutrinos expected
under hadronic models for the TeV UnID sources can be accomodated
within the flux of events observed by IceCube. In hadronic models,
protons are accelerated by some mechanism and produce pions via $pp$
or $p\gamma$ interactions.  A neutral pion decays into two gamma rays
with particle energies of $\approx$25\% (and total flux of
$\approx$50\%) that of the parent proton.  Thus, the luminosity of
non-thermal protons in the TeV UnID sources must satisfy
\begin{eqnarray}
\label{eq:unID}
   \min[1,\tau] \times L_p & \sim & 2 \times 4 \pi d^2 \,
   \varepsilon_{\gamma} \, F_{\varepsilon_{\gamma}} \, N_{\rm UnID}
   \nonumber \\
   ~~ & \sim & 2 \times 10^{35\mbox{--}36} (d/{10 \rm \ kpc})^2 (N_{\rm
     UnID}/10) \rm \ erg \ s^{-1} \ galaxy^{-1}, 
\end{eqnarray}
at $\varepsilon_p \sim \mbox{few TeV}$.  Here, $d$ is the typical
distance to the TeV UnID sources, and $\tau$ is the effective
optical depth for $pp$ or $p\gamma$ interactions.  One can see that
the required luminosity is much smaller than that of supernovae,
$L_{p, {\rm SN}} \sim 10^{41} \rm \ erg \ s^{-1} \ galaxy^{-1}$, for
$\tau \lesssim 10^{-4}$, which indicates that the sources are
significantly more rare.  The observed TeV gamma-ray spectra from TeV
UnID sources can be explained if the proton spectrum extends up to
$E_{\rm max}\simgt 10$ to 100~TeV.

Charged pions, on the other hand, ultimately decay into $e^\pm$ and
muon and electron neutrinos.  Resulting neutrino particle energies are
typically $\approx$5\% (and total neutrino flux, $\approx$15\%) that
of the parent proton.  Thus, the anticipated TeV UnID neutrino flux is
estimated as $\varepsilon_{\nu} F_{\varepsilon_\nu} \sim (3/10)
\varepsilon_{\gamma} F_{\varepsilon_\gamma} \sim 3 \times
10^{-(11\mbox{--}12)}$\,\cgsflux\ at $\varepsilon_{\nu} \sim 1$\,TeV.
Note that roughly equal amounts of energy are delivered to neutral and
charged pions for high-energy parent protons. If the parent proton
spectrum extends up to higher energies with a similar power-law index
of $\Gamma\approx 2.2$, the anticipated neutrino flux at
$\varepsilon_{\nu} \sim 100$\,TeV is $\varepsilon_{\nu}
F_{\varepsilon_\nu} \sim 0.9 \times 10^{-(11\mbox{--}12)}$\,\cgsflux,
consistent with constraints on a Galactic component of the IceCube
Sub-PeV neutrinos, given that less than four of them can be associated
with the TeV UnID sources, as we show below (\S\ref{sec:skydist}).  In
the context of our scenario, an isotropic background of Sub-PeV
neutrinos from the cosmological population of hypernova remnants in
star-forming galaxies at redshifts $z\approx 1$ to 2 is anticipated,
as discussed by \citet{hfl+13}, and addressed by us in
\S\ref{sec:discuss}; the possible spectral break at $\enu\approx
1.5$\,PeV implies acceleration of protons up to $\varepsilon_{p, {\rm
    max}} \simgt 10$\,PeV in this case.  As for any Galactic source
population, however, the cosmic background will not dominate the
Galactic foreground without very strong effects providing either local
suppression of sources or strong cosmic evolution.


Compared to ordinary supernovae, hypernovae exhibit greater explosion
energies, $E_{\rm HN} = 10^{52} E_{52}$\,erg with $E_{52}\sim 1$, but
have a substantially reduced event rate, $R_{\rm HN} \sim 10^{-4}$\,
yr$^{-1}$ galaxy$^{-1}$.  Given that hypernova remnants will have
similar ages as old supernova remnants, $t = 10^5\, t_5$\,yr with
$t_5\sim 1$, the anticipated number in the Galaxy is $R_{\rm HN} \, t
\sim 10$, consistent with them constituting a significant fraction of
the TeV UnID sources. Following the arguments of\citet{Ioka+10unid},
we consider proton acceleration at the forward shock and inelastic
$pp$ interactions as the production process of the highest-energy
gamma-rays and neutrinos. The non-thermal proton luminosity from
hypernova remnants can be estimated as
\begin{equation}\label{eq:L_hn}
  L_{p, {\rm HN}} \sim \epsilon_p \, E_{\rm HN} \, R_{\rm HN} 
    \sim 3 \times 10^{39} \, \epsilon_{p, -1} \, E_{52} \rm \ erg \ s^{-1}
    \ galaxy^{-1},   
\end{equation} 
where $\epsilon_p$ is the efficiency of proton acceleration.  After
escaping the acceleration region, the non-thermal protons interact
with ambient material via inelastic $pp$ collisions.  Since hypernova
remnants can be expected to occur in relatively high-density regions
having $n = 100 \ n_2$\,\percmcube\ with $n_2\sim 1$, the $pp$
interactions may predominantly occur around the hypernova shock.

In the late phase, the radius of the shock is given as $r_{\rm s} \sim
(6.5 \times 10^{19} \mbox{ cm}) \ E_{52}{}^{1/5} \, n_2{}^{-1/5} \,
t_5{}^{2/5}$.  Then, the effective $pp$ optical depth can be estimated
as
\begin{equation}\label{eq3}\label{eq:tau_hn}
  \tau_{pp} \sim n \sigma_{pp} r_{\rm s} \sim 2 \times 10^{-4}.
\end{equation}  
Here $\sigma_{pp} \approx 3 \times 10^{-26}$\,\percmsq\ is the cross
section for the inelastic $pp$ interaction.  This is a lower limit,
since most $pp$ interactions will occur outside of $r_{\rm s}$, where
the weaker magnetic field leads to a low synchrotron brightness from
secondary $e^\pm$.

From Eqs. (\ref{eq:unID}), (\ref{eq:L_hn}), (\ref{eq:tau_hn}), and the
arguments in the previous section, hypernova remnants can be a viable
explanation for both the TeV UnID sources and the Sub-PeV neutrinos if
protons can be accelerated up to $\varepsilon_{p, {\rm max}}\simgt
10$\,PeV.  The angular size of such hypernova remnants can be
estimated as $\theta_{\rm HN} \sim r_s/d \sim 0.2\arcdeg \; (d/10 {\rm
  \ kpc}) \, E_{52}{}^{1/5} \, n_2{}^{-1/5} \, t_5{}^{2/5}$,
consistent with the sizes of the resolved TeV UnID sources.

Now let us estimate the typical maximum energy of protons from
hypernova remnants.  The velocity of the forward shock can be
estimated as $v_{\rm s} \sim r_{\rm s}/t \sim (2.1 \times 10^7 \mbox{
  cm s}^{-1}) \; E_{52}{}^{1/5} \, n_2{}^{-1/5} \, t_5{}^{-3/5}$, and
the magnetic field strength at the shock can be estimated as $B_{\rm
  s} \sim (8 \pi \epsilon_B n m_{\rm p} v_{\rm s}{}^2)^{1/2} \sim (6.0
\times 10^{-4} \mbox{ G}) \; \epsilon_{B,-1}{}^{1/2} \, E_{52}{}^{1/5}
\, n_2{}^{3/10} \, t_5{}^{-3/5}$, where $\epsilon_B =
0.1\epsilon_{B,-1}$ is the efficiency of the magnetic-field
amplification.  The maximum possible energy is given by $E_{p, {\rm
    max}} \sim e B_{\rm s} r_{\rm s} v_{\rm s}/c$, or
\begin{equation}
  E_{p, {\rm max}} \sim (10 \mbox{ PeV}) \; \epsilon_{\rm B, -1}{}^{1/2} \,
    E_{52}{}^{3/5} \, n_2{}^{-1/10} \, t_5{}^{-4/5}, 
\end{equation}
and the corresponding neutrino energy is 
\begin{equation}
  E_{\nu, {\rm max}} \sim 0.05 \, E_{p, {\rm max}} \sim (0.5 \mbox{ PeV})
  \; \epsilon_{B, -1}{}^{1/2} \, E_{52}{}^{3/5} \, n_2{}^{-1/10} \,
  t_5{}^{-4/5}. 
\end{equation}
The above estimate is slightly lower than the suggested 1.5~PeV cutoff
energy of the IceCube Sub-PeV neutrinos (although see \citealt{He+13},
who suggest a cutoff at 0.25~PeV); however, since the Galactic
component provides a subdominant contribution to the total flux, this
is not a critical concern at present. We also acknowledge substantial
observational and theoretical uncertainties, including the calibration
of neutrino energies, and the unaccounted-for contributions from
younger or more energetic remnants. Given these effects, and the
neutrino fluxes calculated for individual sources in
\S\ref{sec:skydist} below, we conclude that hypernova remnants could
be the dominant explanation for the TeV UnID sources and for the
possible Galactic component of Sub-PeV neutrinos.


\section{Expected Sky Distribution and Event Attribution}
\label{sec:skydist}

In order to explore a TeV UnID source origin for a subset of the
IceCube Sub-PeV neutrinos, we have carried out a cross-correlation
analysis of the TeV UnID source population with the arrival directions
of the 28 Sub-PeV neutrinos.

We retrieved the names and positions of 28 TeV UnID or TeV Dark
sources from the online TeVCat of Wakely \& Horan\footnote{TeVCat
  website: \url{http://tevcat.uchicago.edu/}}, and reviewed the latest
publication(s) for each source to determine their TeV gamma-ray
spectral properties and predict possible Sub-PeV neutrino fluxes on
this basis. In the course of this review, we found some source
duplication and mistyping, and made the following changes to the
source list, reducing it to 24 sources for our purposes.

First, the ``Milagro Diffuse'' source in TeVCat has been resolved into
multiple sources by imaging atmospheric \cerenkov\ facilities, and we
choose to treat these sources individually rather than as a
collective. Second, the source TeV~J2032+4130 is duplicated as
MGRO~J2031+41; we list the properties of this source under the latter
designation. Third, VER~J2019+407 has been identified with part of the
shell of SNR~G78.2+2.1 in a recent VERITAS publication
\citep{ver13_2019+407}. Fourth, MAGIC~J0223+403, an unresolved source,
is likely associated with the 3C66B radio galaxy
\citep{magic09_0223+403}, and hence not relevant to the present study.

We present the properties of the known Galactic TeV Unidentified
sources in Table~\ref{tab:unid}. We provide source coordinates,
approximate source size (if resolved at TeV energies), differential
photon flux and power-law photon index at $E_\gamma=1$\,TeV, and
high-energy exponential cutoff energy if known. We also give the name
of the characterizing facility and the reference for the source size
and spectral information. Since we find two discrepant source spectra
for MGRO~J2031+41, we report both values; we adopt the ARGO-YBJ
spectrum \citep{argo12_cygnus} in the analysis that follows, while
noting that the Milagro spectrum \citep{milagro12_cygnus} yields
neutrino predictions that are lower by a factor of two.

Only a fraction of the full sky has been imaged to sufficient
sensitivity to detect the typical TeV UnID source from
Table~\ref{tab:unid}. Nonetheless, the present source list should
offer a fair representation of our Galaxy's TeV UnID source
population, as the Galactic plane survey of HESS \citep{hess11_plane}
and Cygnus survey of VERITAS \citep{ver11_cygnus} have mapped out the
most likely regions for these sources. In the Northern sky --
admittedly, not the most critical in the present context -- wide-area
surveys by the non-imaging facilities Milagro
\citep{milagro_north,milagro_point} and ARGO-YBJ \citep{argo12_cygnus}
exclude the presence of bright uncataloged off-plane TeV sources, as
well. 

For each of the TeV UnID sources, we calculate the neutrino flux under
the hadronic scenario based on its observed TeV gamma-ray spectrum
\citep[e.g.,][]{Kistler_Beacom_2006, Beacom_Kistler_2007}.  We assume
that the neutrino spectrum has the same spectral index as the
gamma-ray spectrum and extends to $E_\nu\simgt 1$\,PeV if a cutoff is
not observed in TeV gamma rays.  Since all current sub-PeV neutrinos
possibly associated with the Galactic plane are cascade events, we
estimate the expected number of cascade neutrino interactions over
$\mbox{30 TeV} < E_\nu < \mbox{1 PeV}$ from each TeV UnID source,
following the optimistic approach outlined in \citet{Laha_et_al_2013};
this is expected to yield a reasonable upper limit.  The resulting
number of expected cascade events in IceCube for an observation time
of $T = 662$ days and fiducial volume of 400~Mton, appropriate for the
present ``contained vertex'' search \citep{Ic3+13-search}, is reported
for each source as the $n_{\nu,i}$ column in Table~\ref{tab:unid}. For
Northern-hemisphere sources, we also present the neutrino flux prior
to accounting for Earth opacity effects, $n^\prime_{\nu,i}$.

We determine the 28 individual Sub-PeV neutrino event types, arrival
directions, and uncertainties by reviewing the publicly-available
slides describing the discovery of the IceCube Sub-PeV cosmic
neutrinos\footnote{Slides for the IPA 2013 meeting may be retrieved
  from the meeting website,
  \url{https://events.icecube.wisc.edu/conferenceTimeTable.py?confId=46}}
\citep{Ic3+13-search,Ic3+13-subpev} and the team's preliminary spatial
clustering analysis \citep{Ic3+13-spatial}. This yields event types
(track or cascade) and arrival directions with sufficient (sub-degree)
precision for our purposes.

Localization uncertainties for cascade-type events are substantial,
vary on an event by event basis with the deposited energy in the
detector, and will have a significant effect on our analysis. We do
not have access to individual event energies or uncertainties;
however, collaboration slides addressing localization uncertainties
\citep{Ic3+13-search} suggest that for a cascade with the median
energy of the Sub-PeV sample, 80~TeV, the 80\%-confidence radius is
26\arcdeg.  Since 80\%-confidence corresponds to 1.79$\sigma$ for a
two-dimensional Gaussian distribution, we adopt a Gaussian with
$\sigma_c = 14.5\arcdeg$ as our localization PDF for each cascade-type
event. This PDF is consistent with a visual inspection of the
likelihood map presented in the team's spatial clustering analysis
\citep{Ic3+13-spatial}.

Similarly, we adopt a Gaussian with $\sigma_t = 2\arcdeg$ as our
localization PDF for track-type events, consistent with the known
localization precision of IceCube for track-type events. Since none of
the seven track-type events lie within 13\arcdeg\ of a known TeV UnID
source, the details of the localization PDF for track-type events has
no effect on our subsequent analysis -- all track events must be
attributed to backgrounds.

At the observed Sub-PeV energies, the sensitivity of IceCube is a
function of declination owing to Earth opacity. For the TeV UnID
sources, we account for these effects using the observed form of each
source's high-energy spectrum and incorporating Earth opacity into the
calculation of $n_\nu$ (Table~\ref{tab:unid}). The isotropic cosmic
background is also affected by Earth opacity, given the observed
average spectrum, as illustrated in a slide from the IceCube discovery
presentation \citep{Ic3+13-subpev}. Specifically, while sensitivity is
roughly uniform over the Southern sky, it drops exponentially in
$\sin(\delta)$ over the Northern sky due to absorption of high-energy
neutrinos by the Earth. Analyzing their presentation of this effect,
the scale length in $\sin(\delta)$ that we infer is 0.66; hence, the
experiment achieves 22\% of its Southern sky sensitivity at the North
celestial pole.

Finally, among the Sub-PeV neutrinos are an estimated $12.1\pm 3.4$
events due to cosmic ray showers in the atmosphere. Given the nature
of the ``contained vertex'' search for Sub-PeV events, which uses the
surface ``IceTop'' array \citep{icetop13} and the outer layers of
IceCube to veto events, the distribution of atmospheric particles that
elude these vetoes is a complex function of declination (zenith
angle). We fix the number of atmospheric events at 12.1 and adopt the
IceCube team characterization of this background from
\citet{Ic3+13-subpev}.

Using the above inputs we model the probability density over the sky
for IceCube Sub-PeV as the sum of three contributions. First, we adopt
the distribution of unvetoed atmospheric events mentioned above for
the 12.1~expected such events. Of the $\approx$16~neutrinos of
non-atmospheric origin, $n_\nu$ are attributed to the TeV UnID
sources, and the remainder are attributed to the isotropic background,
modulated by the decrease in sensitivity over the Northern sky.  Each
TeV UnID source contributes in proportion to its calculated Sub-PeV
cascade neutrino flux at IceCube, $n_{\nu,i}$. Since no track-type
events have positions consistent with any known TeV UnID
source\footnote{We note that the background of atmospheric-induced
  muon neutrinos exceeds that of electron neutrinos by more than an
  order of magnitude at these energies \citep{Laha_et_al_2013}.}, the
TeV UnID probability density map is obtained by convolving the source
contributions with a Gaussian with $\sigma_c=14.5\arcdeg$.

The resulting map of the TeV UnID source-associated probability
density for Sub-PeV neutrinos is presented as the grayscale image in
Fig.~\ref{fig:skymap}. This figure also plots the positions of the
known TeV UnID sources (green circles), with plot symbols scaled
according to the square root of $n^\prime_{\nu,i}$ for each source;
and the positions of the 28 detected Sub-PeV neutrinos, plotted as a
purple $+$ for cascade-type events and as an orange $\times$ for
track-type events.

We calculate the joint likelihood of the IceCube Sub-PeV neutrino
positions as a function of the number of events $n_\nu$ associated
with the TeV UnID source distribution in a one-dimensional fit. We
find a maximum likelihood at $n_\nu = 1.02$ which is preferred over
the $n_\nu = 0$ null hypothesis by $\Delta\log{\cal L} = 0.23$, for an
odds ratio of 1.26:1, without applying any ``Occam's razor'' model
selection penalty for the added parameter of the TeV UnID model. Since
such penalties typically amount to $\Delta\log{\cal L}\approx 1$ or
more, the evidence at present does not rise to a level that would
justify rejecting the null hypothesis, and we do not claim that any of
the known Sub-PeV neutrinos originate in any of the known TeV UnID
sources.

However, if our alternative hypothesis is valid, then our likelihood
function for $n_\nu$ can be normalized and interpreted in a Bayesian
sense as the posterior probability distribution for $n_\nu$, assuming
a uniform prior over the domain of interest. A plot of this posterior
probability distribution for $n_\nu$ is provided in
Fig.~\ref{fig:tevfit}. In this sense, we find that the posterior
probability for $n_\nu \ge 1$ is 71\%, and that the maximum number of
TeV UnID source-associated neutrinos is $n_\nu < 3.8$ at
90\%-confidence. We note that the total number of neutrino cascades
predicted for our optimistic scenario (as provided in the $n_\nu$
column in Table~\ref{tab:unid}) is $n_\nu = \sum_i n_{\nu,i} = 1.18$,
compatible with the current observational limit.


\section{Discussion}
\label{sec:discuss}

We have explored whether, and to what extent, the TeV gamma-ray
unidentified (TeV UnID) sources within the Galaxy might be the source
of some of the 28 Sub-PeV neutrinos recently reported by IceCube
\citep{Ic3+13-subpev}. We reviewed reasoning that suggests some or all
TeV UnID sources may be relatively old ($T\sim 10^5$\,yr) remnants of
hypernova explosions in our Galaxy, and calculated the expected
Sub-PeV neutrino emissions of these remnants. We found that the
expected flux and spectrum of these neutrino emissions are consistent
with their providing a subdominant, Galactic contribution to the total
flux of Sub-PeV cosmic neutrinos.

\begin{figure}
\centerline{\includegraphics[width=5in]{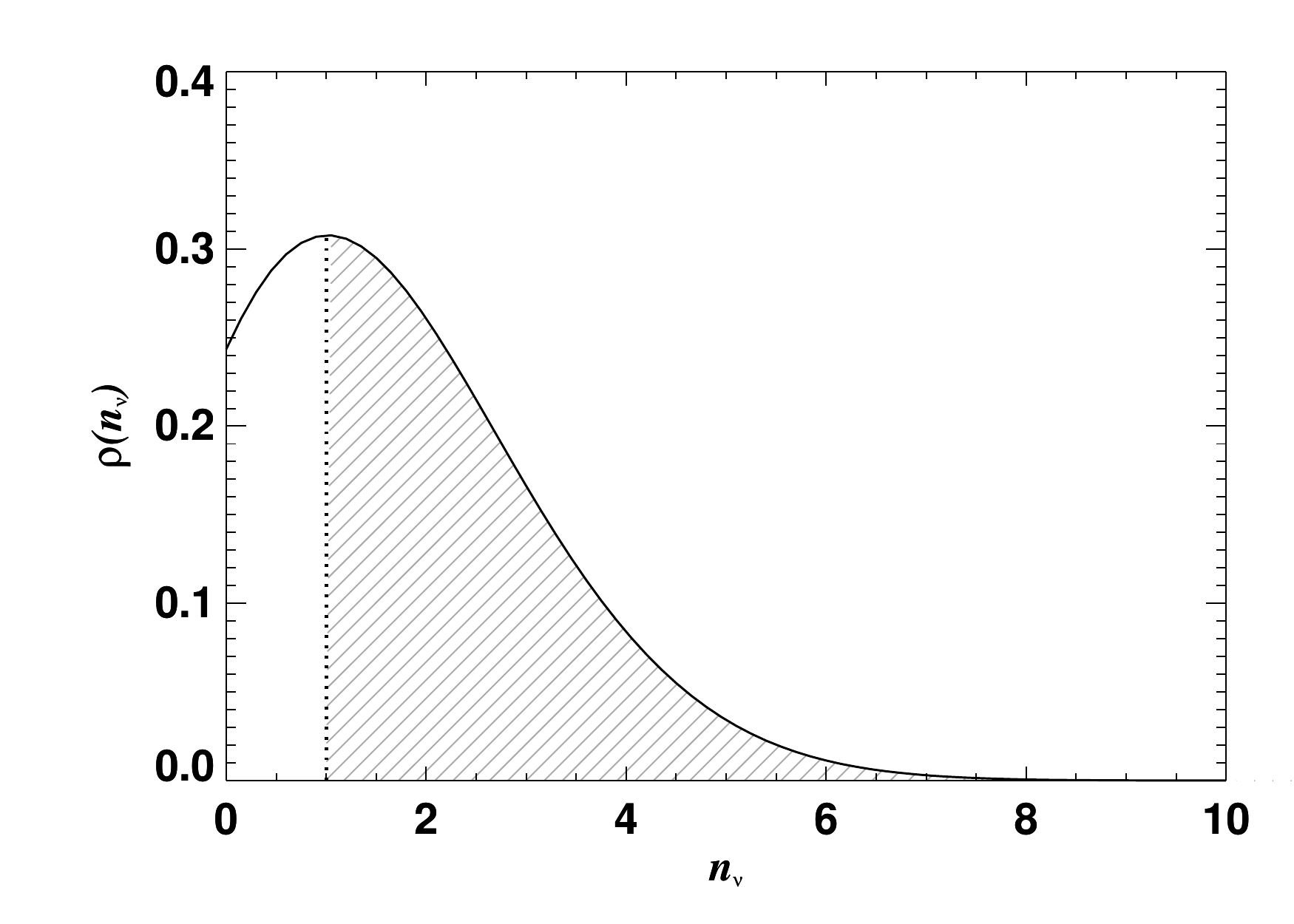}}
\bigskip
\caption[]{%
Posterior probability distribution for $n_\nu$, the number of Sub-PeV
neutrinos distributed on the sky as the flux of TeV UnID sources,
given our assumptions (see text for details). We use a uniform prior
so that this distribution is equivalent to the normalized likelihood
for our single-parameter model. Assuming the validity of our model, we
find that the maximum-likelihood number of associated neutrinos within
the current 28-neutrino sample is $n_\nu=1.02$, and that $n_\nu \ge 1$
with 71\%-confidence (shaded region). We also find that $n_\nu < 3.8$
at 90\%-confidence.
\label{fig:tevfit}}
\end{figure}

We used the positions and TeV gamma-ray fluxes of 24 known TeV UnID
sources to construct a probability density map for Sub-PeV neutrinos
emitted by these sources, and carried out a maximum-likelihood
analysis to explore how many of the Sub-PeV neutrinos may be
attributed to this distribution, rather than the dominant and
quasi-isotropic background. We find that an association of the Sub-PeV
neutrinos with the TeV UnID sources cannot be demonstrated at this
time. However, assuming the validity of our model we find that the
maximum-likelihood number of associated neutrinos within the current
28-neutrino sample is $n_\nu=1.02$, with $n_\nu \ge 1$ at
71\%-confidence and $n_\nu < 3.8$ at 90\%-confidence
(Fig.~\ref{fig:tevfit}). This result is consistent with hadronic model
predictions of the TeV UnID sources, which give an optimistic bound of
$n_\nu\simlt 1.2$.

\begin{figure}
\centerline{\includegraphics[width=7.0in]{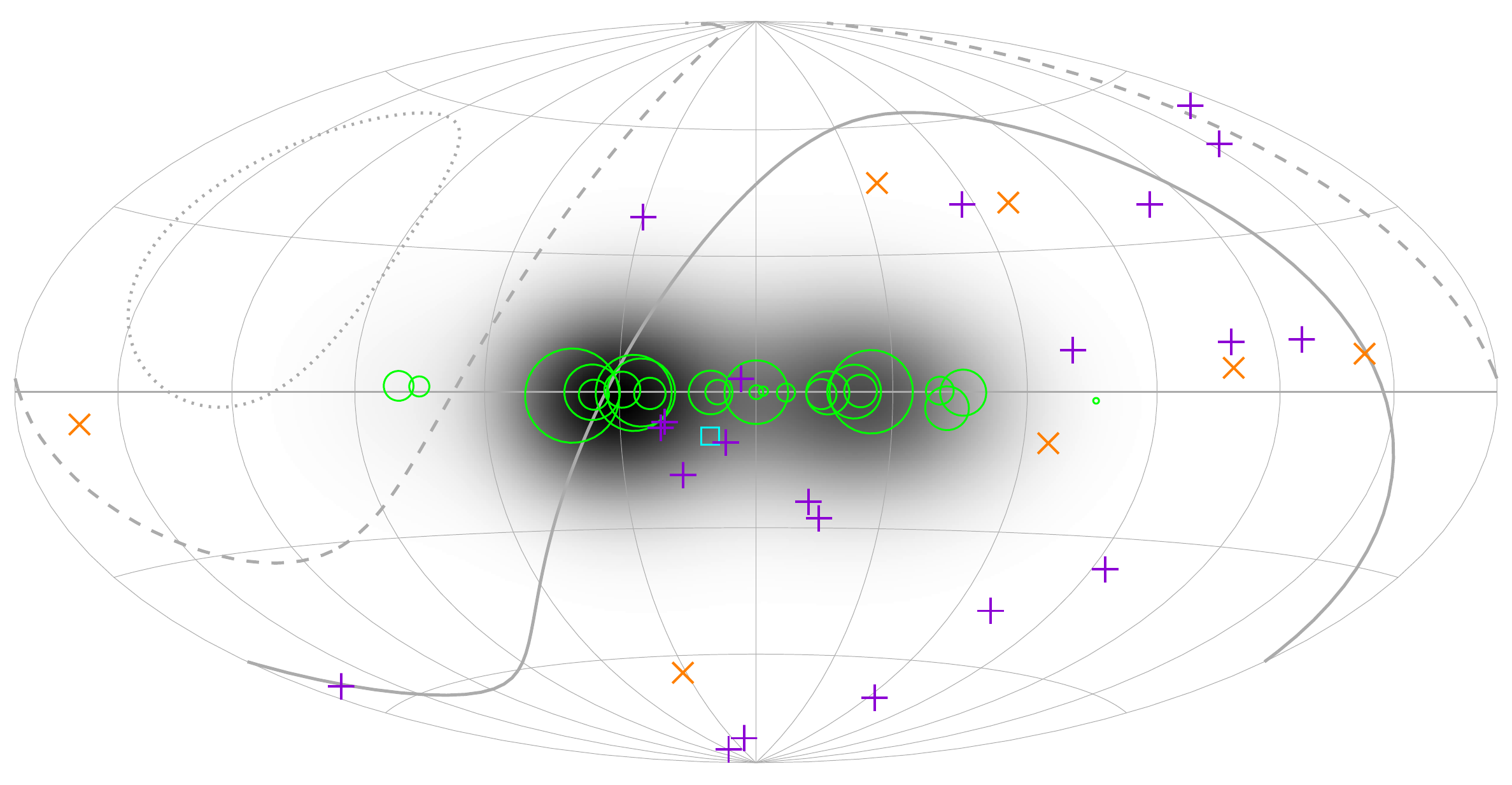}}
\bigskip
\caption[]{%
Sky map in Galactic coordinates of 24 known TeV unidentified sources
(green circles; see Table~\ref{tab:unid}) and 28 IceCube Sub-PeV
neutrino positions (orange $\times$ for track-type events; purple $+$
for cascade-type events).
The greyscale represents the calculated probability density map for
Sub-PeV cascade-type neutrino events which would be associated with
TeV UnID sources.  Symbols for the TeV UnID sources are scaled
according to the square root of the estimated source flux in Sub-PeV
neutrinos, $n^\prime_{\nu,i}$ in Table~\ref{tab:unid}.  The greyscale
map is derived from the source positions and calculated IceCube
Sub-PeV neutrino fluxes $n_{\nu,i}$ (including Earth opacity effects)
by convolution with a Gaussian with $\sigma_c=14.5\arcdeg$,
appropriate for cascade-type events.  It is possible that one or more
of the cascade-type events are associated with the TeV sources, and we
show this probability distribution in Fig.~\ref{fig:tevfit}.  None of
the observed track-type events have a position statistically
consistent with the TeV UnID sources.  The location of the
most-significant point-source position reported by the IceCube team in
their preliminary spatial analysis \citep{Ic3+13-spatial} is indicated
(cyan square).
IceCube sensitivity for Sub-PeV cosmic neutrinos is roughly uniform
over the Southern celestial hemisphere and decreases with increasing
declination over the Northern hemisphere (gray lines at the Equator or
zero declination, and at declinations $+30\arcdeg$ and $+60\arcdeg$)
as an exponential in $\sin\delta$ owing to Earth absorption; see text
for details.
\label{fig:skymap}}
\end{figure}

If our model is valid, we can anticipate detection of further Sub-PeV
neutrinos along the Galactic plane within $|\ell| \simlt 30\arcdeg$ of
the Galactic center and near the Cygnus region at $\ell \approx
80\arcdeg$ (Fig.~\ref{fig:skymap}). While IceCube offers the
highest-sensitivity coverage of the Galactic center, the hotspot in
Cygnus is at northern declinations where ANTARES and other northern
neutrino facilities exhibit greater relative sensitivity.

Given the mixing of neutrino flavors from their production sites
($d\sim 10$\,kpc distant) to Earth, and the roughly 2:1 ratio of
cross-sections for charged vs.\ neutral current neutrino interactions
with bulk matter, we anticipate roughly 2/9 of Sub-PeV neutrinos from
the TeV UnID sources to result in track-type events in
IceCube. Eventually, one or more track-type events must point back to
a TeV UnID source if our model is valid. Given the existence of one
years' worth of as-yet unreported data, the association of 1.0 to 3.8
of the current sample of events with TeV UnID sources yields an
expectation of at least one track event with a further 4 to 18 years'
exposure with IceCube (90\%-confidence upper limit).

Our scenario can also be tested with further TeV gamma-ray
observations by ground-based facilities including HAWC
\citep{DeYoung_et_al_2012} and the CTA \citep{Actis_et_al_2011}, since
we predict the gamma-ray spectra of any TeV UnID sources that are
generating Sub-PeV neutrinos will extend up to very high
energies. Radio and \xray\ observations can also continue to test
leptonic and hadronic models for the TeV UnID sources on a case by
case basis.

Finally, we note that if hypernova remnants are significant sources of
Sub-PeV cosmic neutrinos, their production has been roughly
proportional to the cosmic star formation rate and peaked at redshifts
$z \approx 1$ to 2. Hence, in addition to the Galactic component
associated with TeV UnID sources, we anticipate a diffuse
extragalactic component with a $\Gamma\simgt 2$ power-law spectrum
extending up to a break energy of $\approx$1~PeV, very like the cosmic
neutrino spectrum observed by IceCube. The existence and properties of
this diffuse Sub-PeV neutrino background have been addressed by
\citet{hfl+13}; however, these authors considered only hypernova
remnants in ultraluminous infrared galaxies, and found a diffuse
Sub-PeV neutrino flux that was subdominant to backgrounds from
gamma-ray bursts and active galactic nuclei. Indeed, the dominance of
a quasi-isotropic background over Galactic sources of Sub-PeV
neutrinos suggests that the dominant sources at cosmic distances are
not represented within our Galaxy, as otherwise the Galactic
contribution would dominate by orders of magnitude. However, as an
alternative possibility, we point out that strong cosmic evolution or
suppression of the recent Galactic hypernova rate -- for example, by
our Galaxy's high metallicity
\citep{2006AcA....56..333S,2008AJ....135.1136M} -- will act to
mitigate these effects, and enhance the prominence of the cosmic
background over our Galaxy's contribution. If such effects are strong
enough, then hypernova remnants may even contribute a significant or
dominant portion of the diffuse Sub-PeV cosmic neutrino background
that has now been observed by IceCube.


\acknowledgements

The authors acknowledge the critical reading, comments, and
suggestions of the referee, J.~F. Beacom, which have substantially
improved the manuscript. We also acknowledge useful discussions with
R.~E. Rutledge, A.~D.  Falcone, P. Veres, \mbox{X.-W.} Liu,
\mbox{B.-B.} Zhang, \mbox{X.-H.}  Zhao, D. Cowen, and T. DeYoung.
This work has been supported by a JSPS fellowship for research abroad
(K.K.) and by NASA NNX13AH50G (P.M.).



\begin{deluxetable}{lcrccccccc}

  \tablewidth{0pt}
  \tabletypesize{\scriptsize}
  \tablecolumns{11}

  \tablecaption{TeV Unidentified Sources\label{tab:unid}}

  \tablehead{
    \colhead{Name}  &
    \colhead{R.A.}  &
    \colhead{Dec.}  &
    \colhead{Size ($R$ or $a,b$)}  &
    \colhead{$\phi_0$} &
    \colhead{$\Gamma$} &
    \colhead{$E_{\rm cut}$} & 
    \colhead{$n^\prime_{\nu,i}$} & 
    \colhead{$n_{\nu,i}$} & 
    \colhead{Facility}
  }

\startdata

HESS~J1018$-$589  & 10:17:45.6  & $-$59:00:00    & 0.3\arcdeg  &  0.7 & 2.9  & \nodata 
    & \nodata & 0.00075 & HESS \\ 

HESS~J1427$-$608  & 14:27:52    & $-$60:51:00    & 0.05\arcdeg & 1.3  & 2.16 & \nodata 
    & \nodata & 0.048 & HESS \\ 

HESS J1503$-$582  & 15:03:38    & $-$58:13:45    & 0.4\arcdeg  & 1.6  & 2.4  & \nodata 
    & \nodata & 0.018 & HESS \\ 

HESS~J1507$-$622  & 15:06:53    & $-$62:21:00    & 0.15\arcdeg & 1.8  & 2.24 & \nodata 
    & \nodata & 0.044 & HESS \\ 

HESS~J1626$-$490  & 16:26:04    & $-$49:05:13    & 0.08\arcdeg & 4.9  & 2.18 & \nodata 
    & \nodata & 0.16 & HESS \\ 

HESS J1634$-$472  & 16:34:57.6  & $-$47:16:12    & 0.11\arcdeg & 2.0  & 2.38 & \nodata 
    & \nodata & 0.024  & HESS \\ 

HESS~J1641$-$463  & 16:41:01.7  & $-$46:18:11    & 0.15\arcdeg &  0.8 & 2.0  & \nodata 
    & \nodata & 0.066 & HESS \\ 

HESS J1702$-$420  & 17:02:44    & $-$42:00:57    & 0.08\arcdeg & 2.5  & 2.31 & \nodata 
    & \nodata & 0.043  & HESS \\ 

HESS J1708$-$410  & 17:08:24    & $-$41:05:24    & 0.05\arcdeg & 1.4  & 2.34 & \nodata 
    & \nodata & 0.021 & HESS \\ 

HESS~J1729$-$345  & 17:29:35    & $-$34:32:22    & 0.12\arcdeg & 0.31 & 2.24 & \nodata 
    & \nodata & 0.0075 & HESS \\ 

HESS~J1741$-$302  & 17:41:00    & $-$30:12:00    & 0.2\arcdeg  &  1.1 & 2.78 & \nodata 
    & \nodata & 0.0020 & HESS \\ 

Galactic Center   & 17:45:39.6  & $-$29:00:22    & \nodata     & 2.55 & 2.10 & 15.7    
    & \nodata &  0   & HESS \\ 

Galactic Ridge    & 17:45:39.6  & $-$29:00:22    & 1.6\arcdeg, 0.6\arcdeg   & 5.0 & 2.29 & \nodata 
    & \nodata & 0.092 & HESS \\ 

HESS J1804$-$216  & 18:04:31.2  & $-$21:41:60    & 0.20\arcdeg & 5.7  & 2.72 & \nodata 
    & \nodata & 0.014 & HESS \\ 

HESS~J1808$-$204  & 18:08:00    & $-$20:24:00    & 0.2\arcdeg, 0.1\arcdeg\tnm{a} & 4.0 & 2.4  & \nodata 
    & \nodata & 0.044 & HESS \\ 

HESS J1834$-$087  & 18:34:45.6  & $-$08:45:36    & 0.09\arcdeg & 2.6  & 2.45 & \nodata 
    & \nodata & 0.022 & HESS \\ 

HESS J1837$-$069  & 18:37:38.4  & $-$06:57:00    & 0.12\arcdeg, 0.05\arcdeg & 5.0 & 2.27 & \nodata 
    & \nodata & 0.11 & HESS \\ 

HESS~J1841$-$055  & 18:40:55    & $-$05:33:00    & 0.41\arcdeg, 0.25\arcdeg & 12.8 & 2.41 & \nodata 
    & \nodata & 0.13 & HESS \\ 

HESS~J1843$-$033  & 18:43:40    & $-$03:22:00    & 0.25\arcdeg & 1.0\tnm{b} & 2.20 & \nodata 
    & \nodata & 0.029 & HESS \\ 

HESS~J1857$+$026  & 18:57:11    & $+$02:40:00    & 0.11\arcdeg, 0.08\arcdeg &  6.1 & 2.39 & \nodata 
    & 0.070 & 0.070 & HESS \\ 

HESS~J1858$+$020  & 18:58:20    & $+$02:05:24    & 0.08\arcdeg, 0.02\arcdeg &  0.6 & 2.17 & \nodata 
    & 0.021 & 0.020 & HESS \\ 

MGRO~J1908$+$06   & 19:07:54    & $+$06:16:07    & 0.5\arcdeg  & 4.14       & 2.10 & \nodata 
    & 0.20 & 0.19  & HESS \\ 

VER~J2016$+$372   & 20:16:00    & $+$37:12:00    & \nodata     &  0.19\tnm{c} & 2.1  & \nodata 
    & 0.0092 & 0.0066 & VERITAS \\ 

MGRO~J2031$+$41   & 20:29:38.4  & $+$41:11:24    & 1.8\arcdeg  & 35.\tnm{d} & 3.22 & \nodata & 0.0088 & 0.0070 & Milagro \\ 
                  &             &                & 0.2\arcdeg  & 14.        & 2.83 & \nodata & 0.020 & 0.015 & ARGO-YBJ \\ 

  \enddata

  \tablecomments{Source positions are given in J2000
    coordinates. Source sizes are as quoted in the source reference,
    and are typically the sigma value for the best-fit circular or
    elliptical Gaussian. Source spectra, $N(E_{\rm TeV}) =
    \mbox{($10^{-12}$\,photons \percm\,\persec\,\pertev)}\, \phi_0\,
    E_{\rm TeV}^{-\Gamma}$, are characterized by the differential
    photon flux $\phi_0$ and power-law photon index $\Gamma$ at
    $E_\gamma=1$\,TeV. An exponential cutoff within the energy range
    accessible to current facilities is observed only for the Galactic
    Center source HESS~J1745$-$290, which has $E_{\rm
      cut}=15.7$\,TeV. The expected number of Sub-PeV neutrinos
    detected in 662~days of IceCube exposure using the
    contained-vertex fiducial volume of 400~Mton with (and for
    Northern sources only, without) Earth opacity effects is provided
    as $n_{\nu,i}$ ($n^\prime_{\nu,i}$); see text for details. }

  \tablenotetext{a}{The source extent quoted in
    \citet{hess12_1808-204} for HESS~J1808$-$204, (0.125\arcmin,
    0.037\arcmin), appears to be mistyped; such a source extent would
    be inconsistent with the paper's description of the source as
    resolved in one dimension, with the source map which shows a
    resolved source, and with the spectral analysis, which models the
    source as resolved with radius $R=0.2\arcdeg$.}

  \tablenotetext{b}{TeV flux and spectral information for
    HESS~J1843$-$033 have not been published. The significance map
    from \citet{hess08_1843-033} indicates a total photon flux at
    least twice that of Kes~75, so we adopt a spectral normalization
    of $\phi_\gamma = 1\times 10^{-12}$ cm$^{-2}$ s$^{-1}$ TeV$^{-1}$
    and a photon index $\Gamma=2.2$ spectrum (without cutoff) for this
    source.}  

  \tablenotetext{c}{The normalization for VER~J2016$+$372 is set to
    give 1\% of the Crab photon flux over $\mbox{1\,TeV} < E <
    \mbox{100 TeV}$, i.e., $1.7\times 10^{-13}$\,photons
    \percm\,\persec, consistent with \citet{ver11_cygnus}.}

  \tablenotetext{d}{The Milagro flux measurement at $E_\gamma=10$\,TeV
    \citep{milagro12_cygnus} has been extrapolated to 1~TeV using the
    source power-law photon index $\Gamma=3.22$.}  

  \tablerefs{HESS J1018$-$589 \citep{hess12_1018-589},
             HESS J1427$-$608 \citep{hess08_unid},
             HESS J1503$-$582 \citep{hess08_1503-582},
             HESS J1507$-$622 \citep{hess11_1507-622}, 
             HESS J1626$-$490 \citep{hess08_unid},
             HESS J1634$-$472 \citep{hess06_innergal},
             HESS J1641$-$463 \citep{hess13_1641-463},
             HESS J1702$-$420 \citep{hess06_innergal},
             HESS J1708$-$410 \citep{hess06_innergal},
             HESS J1729$-$345 \citep{hess11_1729-345},
             HESS J1741$-$302 \citep{hess09_plane}, 
             Galactic Center or HESS~J1745$-$290 \citep{hess09_galcenter},
             Galactic Ridge \citep{hess06_galridge},
             HESS J1804$-$216 \citep{hess06_innergal},
             HESS J1808$-$204 \citep{hess12_1808-204},
             HESS J1834$-$087 \citep{hess06_innergal},
             HESS J1837$-$069 \citep{hess06_innergal},
             HESS J1841$-$055 \citep{hess08_unid},
             HESS J1843$-$033 \citep{hess08_1843-033},
             HESS J1857$+$026 \citep{hess08_unid},
             HESS J1858$+$020 \citep{hess08_unid},
             MGRO J1908$+$06 or HESS~J1908+063 \citep{hess09_1908+063}, 
             VER J2016$+$372 \citep{ver11_cygnus},
             MGRO J2031$+$41 or TeV~J2032+4130 \citep{argo12_cygnus,milagro12_cygnus}.}

\end{deluxetable}

\clearpage


\clearpage


\end{document}